# Note: Nanomechanical characterization of soft materials using a micro-machined nanoforce transducer with a FIB-made pyramidal tip

Z. Li[1, a)], S. Gao[1], U. Brand[1], K. Hiller[2], N. Wollschläger[3], F. Pohlenz[1]

[1]*Phyikalisch-Technische Bundesanstalt, 38116 Braunschweig, Germany*
[2]*Technische Universität Chemnitz, 09126 Chemnitz, Germany*
[3] *Bundesanstalt für Materialforschung und -prüfung, 12205 Berlin, Germany*

The quantitative nanomechanical characterization of soft materials using the nanoindentation technique demands further improvements of the performances of instruments, including their force resolution in particular. A micro-machined silicon nanoforce transducer based upon electrostatic comb-drives featuring a force resolution down to ~ 1 nN is described. At the end of the MEMS transducer's main shaft a pyramidal tip is fabricated using a focused ion beam (FIB) facility. A proof-of-principle setup with this MEMS nanoindenter has been established to measure the mechanical properties of ultra-soft polymers. First measurement results demonstrate that the prototype measurement system is able to quantitatively characterize soft materials with elastic moduli down to a few MPa.

The nanoindentation technique belongs to one of the favorable approaches for nanomechanical characterization of materials in small volumes, mainly due to its less requirements on specimen preparation, relatively simple measurement procedure, and well-standardized data evaluation/interpretation methods[1,2]. With a sharp pyramid-like tip (e.g. Berkovich, Vickers), a nanoindentation instrument is even able to determine the mechanical properties of micro-/nano-structurized specimens. In recent times, it has been noticed that much higher force resolution is necessary for a nanoindentation instrument, when soft materials, e.g. polymers and organic materials, are to be measured [3]. Of course, to decrease the requirement on a nanoindentation instrument's force resolution for soft materials, larger spherical indenter [4-5] (e.g. tip radii from tens to hundreds of μm) can be used. These lead, unfortunately, to the loss of the lateral resolution of the measurement and therefore are no longer applicable for micro-structured specimens.

Having been noticed that atomic force microscopes (AFMs) feature not only high lateral resolution, but also high force resolution, in the last years tremendous efforts using AFM (with glued diamond tips in some cases) to characterize the mechanical properties of soft materials have already been carried out [6-7]. However, the disadvantages of cantilever-based AFM nanomechanical measurements e.g. limited indentation depth and force, nonlinearity in case of large cantilever deflection and difficulties to quantitatively characterize the tip area function of an AFM tip, prevent AFMs from quantitative measurements in material testing.

To characterize the mechanical properties of soft materials with lower uncertainties, especially of micro-structured materials, this paper presents a MEMS-based nanomechanical measurement system, which is aimed to bridge the capabilities of currently available nanoindentation instruments and those of AFMs.

As shown in Fig. 1 the MEMS transducer is developed on basis of an electrostatic comb-drive actuation mechanism[8]. The main shaft of the MEMS transducer is symmetrically suspended by folded springs with the stiffness $k_{MEMS}$ along its movable axis. A set of comb drives with a number of finger pairs are designed to move the main shaft in the direction of the +z-axis (symmetry axis of the transducer). The electrostatic force $F_e$ generated by the comb drives[8] is

$$F_e = N \cdot \varepsilon \cdot \frac{t}{g} \cdot U_z^{+2}, \qquad (1)$$

where $N$ is the number of finger pairs for indentation test, $\varepsilon$ the permittivity in air, $t$ the finger height, $g$ the gap between movable and fixed fingers, and $U_z^+$ the DC drive voltage applied to the comb drives. The actual indentation force applied on the surface of a specimen under test is

$$F_{indent} = F_e - k_{MEMS} \cdot h_{indent}, \qquad (2)$$

where $h_{indent}$ is the surface deformation of the specimen under an indentation force.

Being aware that soft materials generally create stronger adhesion forces, another set of comb drives was implemented in the transducer for the generation of pull-off compensation forces. This force is controlled by the DC signal $U_z^-$.

To sense the in-plane displacement of the transducer, an AC signal with the modulation frequency $f_0$ is added to the signals $U_z^+$ and $U_z^-$ with inversed phase, respectively. The output AC current from the transducer is converted to a voltage signal by an I/U converter and then sent to a Lock-in amplifier for further processing. Careful design of

---

a)Author to whom correspondence should be addressed. Electronic mail: zhi.li@ptb.de.

the dual-directional comb drives within the transducer ensures that the Lock-in output $V_s$ is linearly proportional to the axial movement of the transducer's main shaft.

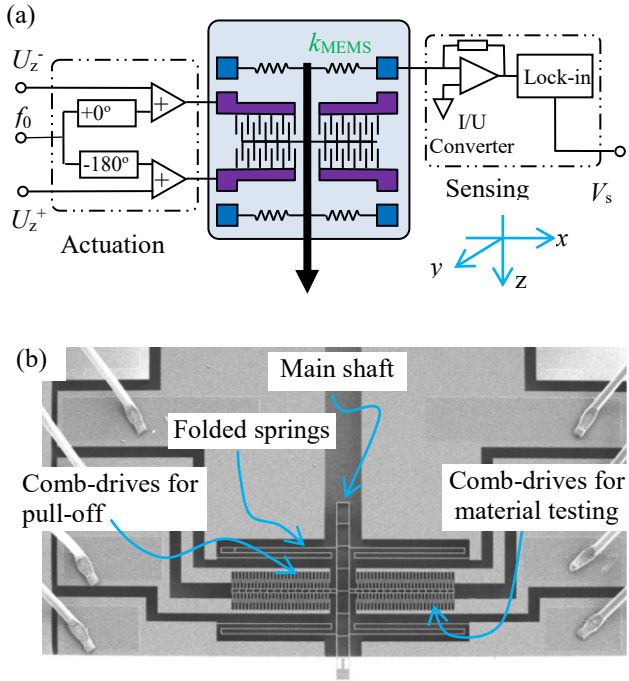

FIG. 1. (a) Schematic of the electrostatic MEMS transducer and the fundamental principle of drive and sensing system for this transducer. (b) SEM image of the MEMS transducer (without pyramidal tip).

The designed MEMS transducer has been fabricated by the silicon Bonding-DRIE technology[9], with a comb finger height $t = 50$ µm and a comb finger gap $g = 3$ µm. One of the prototypes is illustrated in Fig. 1(b). Here the number of finger pairs for indentation/pull-off is $N = 56$, leading to a maximum output electrostatic force $F_e$ up to 40 µN at $U_z^+ = 70$V. With our self-developed stiffness calibration setup[10], the suspending stiffness of the transducer is found to be $k_{MEMS} = 3.6$ N/m ± 5 % (1σ), and the transducer's resonance frequency is about 2.0 kHz. The sensitivity of the capacitive sensing system of the transducer is 0.2 nm[10], indicating a closed-loop indentation force resolution of $\delta F_{indent} = 0.7$ nN.

For the purpose of material testing an indenter with an appropriate shape is necessary at the MEMS transducer's main shaft. Taking into consideration that the MEMS transducer is intended to be used for the characterization of soft materials, whose elastic modulus should be generally far lower than that of silicon, a Vickers-like (i.e. four-sided pyramid) indenter tip is formed directly out of the end of the transducer's main shaft by means of focused ion beam (FIB) fabrication, as shown in Fig. 2. To remove silicon material from the end of the main shaft with FIB, a Gallium ion beam current of about 5 nA at 30 kV has been typically utilized. Subsequently a fine polishing process with 0.5 nA beam current has been used to smooth the surface of the pyramidal tip at the main shaft.

Quantitative nanoindentation tests require careful characterization of the indenter tip in use, especially its projected tip area function (TAF)[2]. The 3D topography of the FIB-shaped pyramidal tip is thereafter measured by a commercial AFM (Dimension icon®, Bruker Corp.[11]). From this measurement the tip area function $A_p$ of the silicon tip within its valid indentation depth is evaluated[12], as listed in Table 1. The semi-angle of the opposite faces of the pyramidal tip amounts to 68.7 ± 0.2°, well close to its ideal value 68°. As can be seen from Fig. 2(c), the FIB fabricated silicon tip has evident tip-rounding, with an equivalent tip radius $R_{rounding} = 3.96$ µm, which leads to a relatively large deviation of $C_0$ from its ideal value (i.e. 24.5).

TABLE I. TAF coefficients of the FIB-shaped pyramidal tip, where $A_p = \sum_{i=0} c_i \cdot h^{2^{1-i}}$.

| $C_0$ | $C_1$ | $C_2$ | $C_3$ | $C_4$ | $C_5$ |
|---|---|---|---|---|---|
| 21.5 | 19081.5 | -17411.4 | 10720.7 | 7604.7 | 7083.8 |

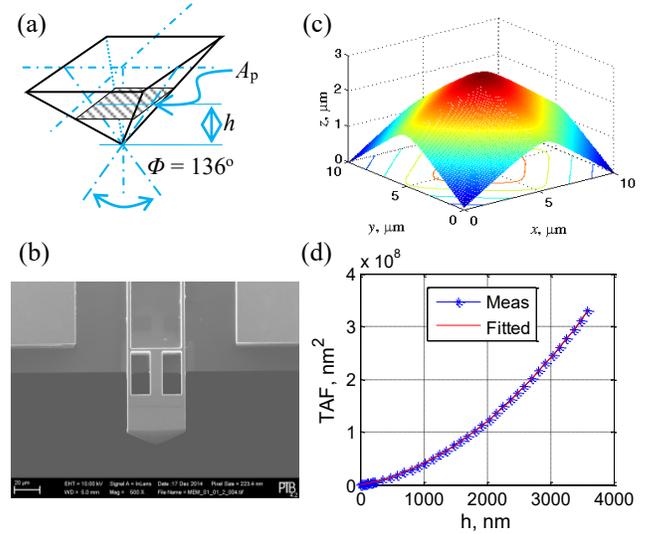

FIG.2 (a) Schematic diagram of a Vickers indenter tip. (b) Electron microscope image of the FIB- shaped Vickers indenter at the end of the main shaft of the MEMS nano-force transducer. (c) 3D topography of the FIB-shaped indenter tip measured by an AFM. (d) Tip area function of the FIB-shaped tip evaluated from the AFM-Image.

To investigate the capabilities of the MEMS nanoindenter, an experimental setup has been established, in which a cost-effective three-axis stage integrated with a 3D closed-loop piezo-positioning system (NanoMAX 311, Thorlabs[13]) is used to engage the MEMS transducer and the sample under test. Owing to the high force resolution of the MEMS nanoindenter, sample engagement with a typical preload of about 30 nN works well. After engagement, quasi-static nanoindentation testing is carried out, where the typical loading, holding and unloading time in use are 10 s, 3 s and 10 s, respectively.

Within the proof-of-principle test, a vulcanized isoprene rubber (IR), which was recently employed as a testing material for AFM nanomechanical measurements[14], has been investigated. One of the typical indentation force-depth curves for this material is shown in Fig. 3. It can be seen that the rubber under test can be treated as an elastic material even for deep indentation. To extract the mechanical properties of IR from the measurement data, the Oliver-Pharr model[1] (i.e. $F = \alpha[h-h_f]^m$) is applied to analyse the unloading curves, as shown in Fig. 3. The residual indentation depth $h_f$ in the experiments tends to 0, and the power law exponents of the IR is experimentally determined to $m = 6.4$. The actual contact depth between the pyramidal indenter and the specimen surface $h_c$ is calculated as $h_c = h_{max} - 0.75 * F_{max}/S$, where $S$ is the slope of the unloading curve at its initial point, as shown in Fig. 3.

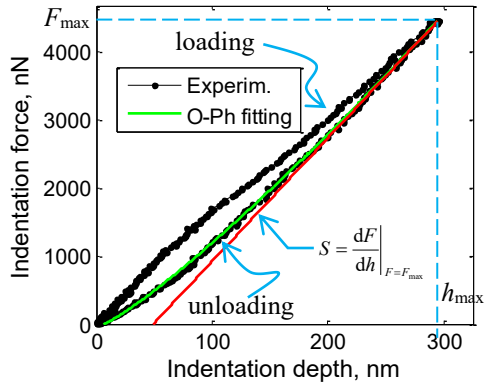

FIG. 3. Typical indentation depth-force curve obtained for the vulcanized IR sample.

A series of indentation tests were carried out on the IR specimen. The indentation modulus $E_{IT}$ of the specimen versus indentation depth is shown in Fig. 4. In the case of $F_{max} = 21.7$ µN, the maximum indentation depth $h_{max}$ amounts to 1.27 µm. It can be seen clearly that the elastic modulus of the IR specimen increases rapidly, especially when the depth $h_c$ becomes less than 300 nm.

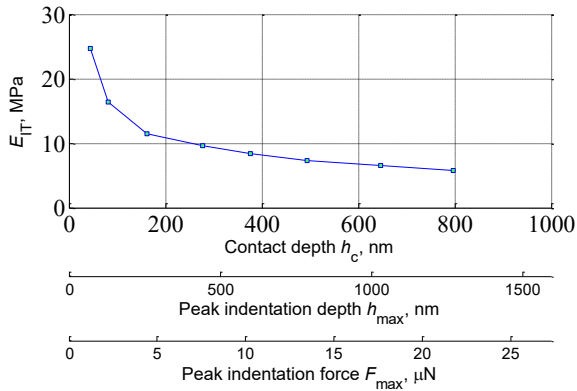

FIG. 4. Depth-dependent indentation modulus of the vulcanized IR sample measured by the MEMS nanoindenter.

For comparison, the IR sample has also been measured using a commercial nanoindentation instrument (Triboscope TI-950, Hysitron Inc.[15]) with a Berkovich indenter. The indentation modulus of the IR sample converges to 2.1±0.2 MPa, with contact depths $h_c > 15$ µm. Within the range of shallow contact depths (i.e. $h_c < 2000$ nm), as shown in Fig. 5, especially for $h_c \geq 300$ nm (or a maximum indentation force for this Berkovich tip of $F_{max} \geq 10$ µN), the measurement results obtained by the commercial instrument and those by the MEMS nanoindenter coincide quite well. However, due to the relative high preload force of the commercial nanoindenter (2 µN), the calculated $h_c$ can have a relatively large (offset) deviation, leading to the evident difference of the measured $E_{IT}$ for $h_c < 300$ nm.

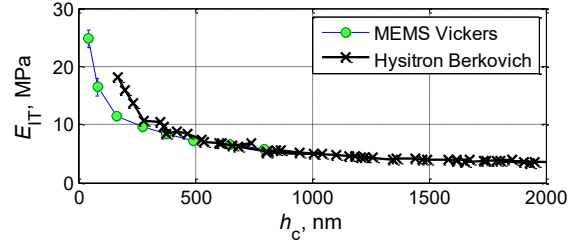

Fig. 5 Comparison between the measurement results of the vulcanized IR obtained by the MEMS Nanoindenter and those by a commercial nanoindentation instrument (Hysitron Triboscope).

To sum up, a micro-machined nanomechanical measurement system has been developed, which utilizes electrostatic comb-drives to generate test forces and to sense the indentation depth simultaneously. The system features a closed-loop force resolution down to 0.7 nN. A Vickers-like indenter tip was formed directly to the MEMS transducer's main shaft by FIB fabrication. Preliminary experiments validate that the MEMS nanoindenter is able to measure the mechanical properties of soft materials with elastic moduli going down to several MPa.

It is therefore believed that this micro-machined nanoindenter can well bridge the gap between nanoindentation instrument and AFM contact-based material testings. Further developments of the MEMS nanoindenter, e.g. extending the test force up to the mN-range, functional extension for dynamic measurements, and integrated fiber interferometer for in-situ indentation depth measurement, are under consideration.

This research is supported by the European Union by funding the European Metrology Research Program (EMRP) project "Traceable measurement of mechanical properties of nano-objects (MechProNo)". The authors would like to thank Prof. K. Nakajima (Tokyo Institute of Technology) for providing the samples of vulcanized IR.